\documentstyle[aps,prl,twocolumn]{revtex} 
\tighten
\begin{document}
\draft
\def\gev{\mathrel{\rm GeV}}
\def\asym{{\rm asym}}
\def\df{{\rm d}}
\def\L{{\rm L}}
\def\P{{\rm P}}
\def\R{{\rm R}}
\def\eq{{\rm eq}}
\def\vev{{\em vev}}
\twocolumn[\hsize\textwidth\columnwidth\hsize\csname@twocolumnfalse\endcsname
\title{Scale of $SU(2)_{\rm R}$ symmetry breaking and leptogenesis}
\author{Ernest Ma$^a$, Subir Sarkar$^b$ and Utpal Sarkar$^c$}
\address{$^a$ \sl Department of Physics, University of California,
          Riverside, California 92521, USA\\
         $^b$ \sl Theoretical Physics, University of Oxford,
          1 Keble Road, Oxford OX1 3NP, UK\\
         $^c$ \sl Physical Research Laboratory, Ahmedabad 380 009, India}
\date{\today}
\maketitle
\begin{abstract}
Models of leptogenesis often invoke the out-of-equilibrium decays of
heavy right-handed neutrinos in order to create a baryon asymmetry of
the universe through the electroweak phase transition. Their presumed
existence argues strongly for the presence of an $SU(2)_\R$ gauge
symmetry. We study the equilibrating effects of the resulting
additional right-handed interactions and find that successful
leptogenesis requires $m_N\gtrsim10^{16}$ GeV if $m_N>m_{W_\R}$,
and $m_{W_\R}\gtrsim2\times10^5\gev(m_N/10^2\gev)^{3/4}$ if
$m_{N}<m_{W_\R}$, where $m_{N}$ is the mass of the lightest
right-handed neutrino. A better bound
$m_{W_\R}\gtrsim3\times10^6\gev(m_N /10^2\gev)^{2/3}$ is obtained if
leptogenesis occurs at $T>m_{W_\R}$.  We show also that the
$m_N>m_{W_\R}$ option is excluded in a supersymmetric theory with
gravitinos.
\end{abstract}
\vglue.4truecm]
\narrowtext 

It is now accepted that neutrinos do have small masses, thus accounting for 
the atmospheric neutrino anomaly \cite{atm} and the solar neutrino puzzle 
\cite{sol}, and perhaps also providing a fraction of the dark matter of the 
universe \cite{dark}.  A natural solution to the smallness of neutrino masses 
is to consider them as Majorana particles. The nonconservation of lepton 
number at some large scale $M$ would induce an effective dimension-5 operator 
\cite{dim5} $h\ell_\L \ell_\L\phi\phi/M$, where $\phi$ is the usual Higgs 
doublet.  As $\phi$ acquires a nonzero vacuum expectation value (\vev),
$\langle\phi\rangle=v$, to break the electroweak gauge symmetry, neutrinos 
obtain small masses as well: $ m_\nu=hv^2/M$.

An additional appeal of this solution is that it offers an elegant
mechanism for generating the baryon asymmetry of the universe.  As is
well known, interactions which violate $B+L$ while conserving $B-L$
are unsuppressed by sphaleron processes at high temperatures
\cite{krs}.  Thus any primordial $L$ asymmetry would be (partly)
converted into a $B$ asymmetry --- a process termed leptogenesis.  The
required lepton asymmetry can be generated in two possible ways
through the lepton number violation responsible for neutrino masses
\cite{fy,ma}.  One can introduce right-handed singlet neutrinos which
acquire large Majorana masses, resulting in small Majorana masses for
the left-handed neutrinos, through the so-called seesaw mechanism
\cite{seesaw}.  Decays of the right-handed neutrinos do not conserve
lepton number, thereby generating a primordial lepton asymmetry
\cite{fy}. Alternatively, one can extend the standard model to include
very heavy Higgs triplet scalars whose couplings break lepton number
explicitly.  They would naturally acquire tiny seesaw {\vev}\,s,
thereby inducing small masses for the left-handed neutrinos and their
decays would also generate the primordial lepton asymmetry \cite{ma}.

In both routes, the scale of lepton number violation $M$ is arbitrary.  
It is thus natural to consider a left-right symmetric model where both 
possibilities are realized and $M$ is related to the left-right symmetry 
breaking scale \cite{lr}. This is a natural framework for explaining 
parity nonconservation at low energies; it may also be embedded in 
interesting grand unified theories.  Within this broader context, an 
important change in the conditions of leptogenesis occurs because the 
right-handed neutrinos must now interact with the $SU(2)_\R$ gauge bosons.

In this Letter, we examine the effect of the interactions of the
right-handed gauge bosons $W_\R$ on the generation of the primordial
lepton asymmetry of the universe.  We conclude that for this to be
phenomenologically successful, $W_\R$ should be heavier than the
lightest right-handed neutrino:
$m_{W_\R}\gtrsim2\times10^5\gev(m_N/10^2\gev)^{3/4}$, unless
$m_N\gtrsim10^{16}\gev$ in which case $m_{W_\R}/m_N\gtrsim0.1$. A
better bound $m_{W_\R}\gtrsim3\times10^6\gev(m_N/10^2\gev)^{2/3}$ is
obtained if leptogenesis occurs at $T>m_{W_\R}$.

In left-right symmetric models the quarks and leptons transform under the 
group $SU(3)_{\rm c}\otimes\,SU(2)_\L\otimes\,SU(2)_\R\otimes\,U(1)_{B-L}$ as
\begin{eqnarray}
 q_{i\L} \equiv \pmatrix{u_{i\L} \cr d_{i\L}} \sim (3, 2, 1, 1/3),
 \nonumber \\ 
 q_{\alpha\R} \equiv \pmatrix{u_{\alpha\R} \cr d_{\alpha\R}} 
 \sim (3, 1, 2, 1/3), \nonumber \\
 \ell_{i\L} \equiv \pmatrix{\nu_{i\L} \cr e_{i\L}} \sim (1, 2, 1, -1),
 \nonumber \\
 \ell_{\alpha\R} \equiv \pmatrix{N_{\alpha\R} \cr e_{\alpha\R}} 
 \sim (1,1,2,-1). 
\end{eqnarray}
The Higgs bidoublet $\phi\equiv(1,2,2,0)$ breaks electroweak symmetry, 
whereas the left-right symmetry is broken by the $SU(2)_\R$ Higgs triplet 
$\Delta_\R\equiv(1,1,3,-2)$. Left-right parity also requires the existence 
of an $SU(2)_\L$ Higgs triplet $\Delta_\L\equiv(1,3,1,-2)$.

To begin with, we shall ignore the effects of the triplets for the
generation of the lepton asymmetry and consider the interactions of
the leptons alone:
\begin{eqnarray}
 {\cal L} &=& f_{i \alpha} \overline{\ell_{i\L}} \ell_{\alpha\R} \phi
 + f_{\L ij} \overline{{\ell_{i\L}}^c} \ell_{j\L} \Delta_\L
 + f_{\R\alpha\beta} \overline{{\ell_{\alpha\R}}^c} \ell_{\beta\R} 
  \Delta_\R  \nonumber \\ 
 && + {1 \over 2} g_\L \overline{\ell_{i\L}} \gamma_\mu \tau^a_{ij} 
 \ell_{j\L} W_\L^{a\mu} + {1 \over 2} g_\R \overline{\ell_{\alpha\R}} 
 \gamma_\mu \tau^a_{\alpha\beta} \ell_{\beta\R} W_\R^{a\mu}.
\end{eqnarray}
Note that $B-L$ is now a {\em local} symmetry, hence it cannot be directly
violated at high energies.  Its violation occurs only when the left-right 
symmetry is broken by the {\vev} $v_\R$ of the right-handed Higgs triplet 
$\Delta_\R$. This gives Majorana masses to the right-handed neutrinos, and 
lepton number is violated in their decays:
\begin{equation}
 N_{\alpha\R} \to \ell_{i\L} + \phi^\dagger~~ {\rm and} ~~ 
 {\ell_{i\L}}^c + \phi.
\end{equation}
If the couplings $f_{i \alpha}$ are complex and if these decays satisfy the 
out-of-equilibrium condition, then they can generate a primordial $B-L$ 
asymmetry. There are two contributions to the magnitude of this asymmetry, 
the first coming from the interference of the tree-level diagrams with the 
vertex diagrams \cite{fy,ver,moh} and the second from that of the former 
with the self-energy diagrams \cite{self}. Although one starts with real and
diagonal masses for the heavy neutrinos, the loop diagrams introduce complex 
phases and hence an amount of $CP$ violation, denoted as $\eta$ in the 
following.

We assume the neutrino masses to be hierarchical ($M_{N_{3\R}}\gg\,M_{N_{2\R}}
\gg\,M_{N_{1\R}}=m_N)$.  It is the decay of the {\em lightest} right-handed 
neutrino which will determine the final lepton asymmetry, hence we shall 
only be concerned with the scale $m_N$.  The lepton asymmetry $n_L\equiv\,
n_l-n_{l^c}$ will evolve according to the transport equation \cite{fry}:
\begin{eqnarray}
 \frac{\df n_L}{\df t} + 3 H n_L &=
 &\eta \Gamma_N [n_N - n_N^{\eq} ] 
 - {1 \over 2} \left(\frac{n_L}{n_\gamma} \right) n_N^{\eq} \Gamma_N
 \\ \nonumber
 &&- 2 n_\gamma n_L \langle \sigma |v| \rangle,   
\label{nl}
\end{eqnarray}
where $\Gamma_N$ is the thermally-averaged decay rate of the
right-handed neutrinos $N_{\alpha\R}$, and $n_N$ their number density,
with the equilibrium value 
\begin{equation}
 n_N^{\eq}=\left\{\begin{array}{ll}
 (45/2\pi^4) s g_*^{-1} , &m_N \ll T, \\
 (45/4\pi^4) s g_*^{-1} \sqrt{\pi/2} (m_N/T)^{3/2} {\rm e}^{-m_N/T},
  &m_N \gg T . \end{array} \right.
\end{equation} 
In the above, $s$ is the entropy density, $n_\gamma$ is the photon
density, and $g_*$ the effective number of interacting relativistic
degrees of freedom. The second term on the left-hand side (lhs) of
Eq.~(\ref{nl}) accounts for the expansion of the universe (where
$H\simeq1.7g_*^{1/2}T^2/M_{\P}$ is the Hubble rate) and the term
$\langle\sigma|v|\rangle$ is the thermally-averaged lepton-number
violating scattering cross section.

Similarly, the density of the heavy particles satisfies the equation
\begin{equation}
 \frac{\df n_N}{\df t} + 3 H n_N = - \Gamma_N (n_N - n_N^{\eq})
 - ({n_N}^2 - {n_N^{\eq}}^2) \langle \sigma_N |v| \rangle.
\label{nN}
\end{equation}
The first term on the right-hand side (rhs) accounts for the decays
(and inverse decays) of the heavy right-handed neutrinos.  The second
term on the rhs, although usually neglected in discussions of
out-of-equilibrium decays, is in fact crucial in the present context.
This is the lepton-number {\em conserving} thermally-averaged
scattering cross section of the right-handed neutrinos
$N_{\alpha\R}$. It is instrumental in initially equilibrating the
number density of the right-handed neutrinos but, as we shall see, it
severely depletes the amount of lepton asymmetry generated by their
decays.

For convenience we define the parameters $K\equiv \Gamma_N/H$ 
and $K_N\equiv2n_\gamma\langle\sigma_N|v|\rangle/ H$ at 
$T=m_N$. For $K\ll\,1$ at $T\sim\,m_N$, the system is far from 
equilibrium, hence the last two terms in Eq.~(\ref{nl}) (the 
ones responsible for the depletion of $n_L$) are negligible.  
In this limit, if $K_N\ll\,1$, the asymptotic solution is 
$n_L^{\asym}/s\simeq\eta/g_*$.  However, for $K_N\gg1$ there is a 
strong suppression of the abundance. For $K_N\sim1$ there is 
already a suppression $n_L\sim0.04n_L^{\asym}$, while in the 
range $1<K_N<10^{3}$ the suppression may be approximated as 
$n_L\sim0.04n_L^{\asym}/K_N$.  Beyond this range the 
suppression is somewhat faster than linear in $K_N$.

We shall now assume conservatively that for an adequate lepton
asymmetry to be generated, in addition to the usual out-of-equilibrium
condition \cite{kt}, viz.  $K\lesssim1$, we also require
$K_N\lesssim1$. The difference is that if the former condition is not
fulfilled, any primordial asymmetry will also be erased.  This does
not happen if the latter condition is not satisfied, since $K_N$
measures the approach to {\em kinetic} rather than chemical
equilibrium.  Nevertheless, a large value of $K_N$ suppresses the
generation of a lepton asymmetry and in the following we show that
$W_\R$ mediated scattering processes are important in this context.

We first consider the case $m_N>m_{W_\R}$.  At the time when $N_{1\R}$
decays, it is still interacting with $W_\R$.  If these interactions
are sufficiently fast, equilibrium of $N_{1\R}$ with the decay
products will be maintained, thus {\em preventing} the generation of
any lepton asymmetry. The requirement for the $SU(2)_\R$ interactions
\begin{equation} 
e^-_\R + W_\R^+ \to N_\R \to e^+_\R + W_\R^- 
\label{scat}
\end{equation}
to fall out of equilibrium is
\begin{equation}
 {g^2_\R \over 8 \pi} {T} \lesssim 1.7 g_*^{1/2} {T^2 \over M_{\P}}
 \qquad {\rm at}~~T = m_N,
\end{equation} 
so that for generating a lepton asymmetry we require
\begin{equation}
 m_N \gtrsim 10^{16}\gev,
\label{lower}
\end{equation} 
where we take $g_\R^2=g_\L^2=0.4$, $g_*\sim10^2$, and
$M_P\sim10^{19}\gev$.  Note that this stringent bound comes from the
fact that the gauge coupling $g_\R$ is of order unity.  In the usual
leptogenesis scenario without $SU(2)_\R$ interactions, the
corresponding coupling is a Yukawa coupling which may be very much
suppressed.  Since $m_{W_\R}/m_N \sim g_\R/f$, where
$f\lesssim\sqrt{4\pi}$ is a reasonable assumption, we conclude that
$m_{W_\R}/m_N\gtrsim0.1$ in this case.

The above lower bound is in conflict, however, with an independent
upper bound on $m_N$ if the theory is supersymmetric and includes a
gravitino, as we discuss below. It is of course necessary that
leptogenesis occurs {\em after} inflation.  In supersymmetric theories
the temperature at the begining of the radiation-dominated era
following inflation is restricted from considerations of the thermal
production of massive gravitinos \cite{grav}.  Since they interact
only gravitationally and thus decay after nucleosynthesis, the
abundances of the synthesized elements can be drastically altered, in
conflict with observations \cite{ens,ab}. This imposes a bound on the
`reheating' temperature at the begining of the radiation-dominated era
which is usually quoted to be of ${\cal O}(10^9)\gev$.  However, a
recent reevaluation of the gravitino production rate \cite{plasma}, in
conjunction with the nucleosynthesis constraints \cite{ab} strengthens
this to $T_{\rm reheat}\lesssim10^{7}\gev$ for weak-scale gravitinos.
Even taking into account that particles of mass as high as
$\sim10^3T_{\rm reheat}$ may be produced with sufficient abundance for
successful leptogenesis \cite{riotto}, this argument severely
restricts the maximum possible mass of the right-handed neutrino:
\begin{equation}
 m_N \lesssim 10^{10}\gev. 
\label{bound}
\end{equation}
Combined with the lower bound of Eq.~(\ref{lower}) on $m_N$, this
rules out the possibility of leptogenesis if $m_N>m_{W_\R}$.  It would
seem that this argument can be evaded if the gravitino is in fact the
lightest supersymmetric particle, and is thus stable.  Even so there
would be a constraint from the requirement that they do not
`overclose' the universe which relaxes the upper bound on the reheat
temperature to $T_{\rm reheat}\lesssim10^{11}\gev$
\cite{ens,plasma}. This requires $m_N\lesssim10^{14}\gev$ for
leptogenesis, so there is still a conflict with the lower bound of
Eq.~({\ref{lower}).

We now discuss the effect of $W_\R$ interactions when $m_N<m_{W_\R}$.
In this case, we must consider both $T=m_N$ and $T=m_{W_\R}$. The
scattering processes
\begin{equation}
e^\pm_\R + N_\R \to W^\pm_\R \to e^\pm_\R + N_R
\end{equation}
are important for bringing the right-handed neutrinos into
equilibrium.  The condition that this reaction departs from
equilibrium when the right-handed neutrinos decay is
\begin{equation}
 {g_\R^4 \over 16 \pi} {T^5 \over m_{W_\R}^4} \lesssim 1.7 g_*^{1/2} 
{T^2 \over  M_{P}} \qquad {\rm at}~~T = m_N,
\end{equation}
which translates into 
\begin{equation}
 m_{W_\R} \gtrsim  2 \times 10^5\gev \left( {m_N \over 10^2\gev} 
\right)^{3 \over 4}. 
\label{bound1}
\end{equation}
Another important process is the scattering of $W_\R$'s into $e_\R$'s 
through $N_\R$ exchange:
\begin{equation}
W^\pm_\R + W^\pm_\R \to e^\pm_\R + e^\pm_\R.
\end{equation}
This is the analog of the standard-model process $W^\pm_\L + W^\pm_\L \to 
e^\pm_\L + e^\pm_\L$ through $\nu_\L$ exchange \cite{utpal}.  The condition 
that this reaction departs from equilibrium is
\begin{equation}
{3g_\R^4 \over 32 \pi} {m_N^2 T^3 \over 
  m_{W_\R}^4} \lesssim 1.7 g_*^{1/2} \frac{T^2}{M_\P} \qquad {\rm at}
  \:\: T = m_{W_\R},
\end{equation}
which translates into
\begin{equation}
 m_{W_\R} \gtrsim 3 \times 10^6\gev \left( {m_N \over 10^2\gev} 
\right)^{2 \over 3}.
\end{equation}

We note that Eq.~(15) applies \cite{new} if leptogenesis occurs at
$T\simeq\,m_N$ and Eq.~(18) applies if it occurs at $T>m_{W_\R}$.
Recognizing that Eq.~(18) is a better bound than Eq.~(15) and noting
that $m_N$ should exceed the electroweak breaking scale of $10^2\gev$,
we have an absolute lower bound of $3 \times 10^6\gev$. This may be
further improved if $m_N$ is of ${\cal O}(10^3)\gev$ \cite{pil}, in
which case $m_{W_\R}\gtrsim10^7\gev$.  A similar bound was mentioned
previously \cite{moh} in the context of having $m_N \sim 10^7\gev$.
Note also that the analog of Eq.~(16) for the neutral $SU(2)_\R$ gauge
boson, {\it i.e.}  $Z_\R+Z_\R\to\,N+N$, is less important because
$m_N$ is heavy.

Consequently, if a right-handed gauge boson is observed with a mass
below $\sim10^7$~GeV, it will necessarily imply that right-handed
neutrinos {\em cannot} have generated the lepton asymmetry of the
universe. Moreover, since the interactions of $W_\R$ would have erased
all primordial $(B-L)$ asymmetry, the observed baryon asymmetry must
have been generated at a scale lower than the $SU(2)_\R$ symmetry
breaking scale $M_\R$.

Finally we discuss the contributions of the Higgs triplet scalars to
leptogenesis.  As mentioned earlier, left-right parity breaks along
with the $SU(2)_\R$ symmetry when the field $\Delta_\R$ acquires a
\vev.  The masses $M_\Delta$ of $\Delta_\L$ and $\Delta_\R$ are
initially equal, but their \vev\,s are rather different, being related
by $v_\L\sim\,v^2/v_\R$.  Beyond the terms given in Eq.~(2), we also
have $\Delta_\L\Delta_\R\phi\phi$.  Since $v_\L$ is very small,
$\Delta_\R$ will dominantly decay into 2 leptons but rarely into 2
scalars, so it cannot create a lepton asymmetry. However, since $v_\R$
is of ${\cal O}(M_\R)$, the decays of $\Delta_\L$ can contribute
significantly to the lepton asymmetry of the universe \cite{ma,lrpat}.
Moreover, since $\Delta_\L$ does not interact with $W_\R$, the
existence of right-handed gauge bosons does not change this
conclusion. The lepton-number conserving interaction,
$\Delta_\L^\dagger + \Delta_\L\to\,W_\L^\dagger+W_\L$, is the most
efficient one at bringing the number density of $\Delta_\L$ into
equilibrium.  So for the lepton asymmetry to survive, this interaction
should be out of equilibrium, implying a lower bound on $M_\Delta$ of
$\sim10^{13}\gev$ \cite{ma}.

As an example, in a realistic supersymmetric SO(10) grand unified
theory, where two 10-plet and one 126-plet contribute to the fermion
masses \cite{bis} and the Majorana Yukawa couplings of the
right-handed neutrinos can be calculated, the left-right symmetry
breaking scale is very high and the Yukawa couplings are also quite
large. In this case, $m_N>m_{W_\R}$ and the bound of Eq.~(\ref{lower})
applies.  As discussed earlier, this means that it will not be
possible to create enough right-handed neutrinos after reheating
without also creating an unacceptable abundance of gravitinos.
However it may well be possible to have a scalar potential which
allows $M_\Delta\sim10^{13}$~GeV, so that decays of the Higgs triplets
can generate a lepton asymmetry leading to successful baryogenesis
through the electroweak phase transition.

To conclude, we have shown that $W_\R$ interactions will in general
bring the number density of the right-handed neutrinos into
equilibrium.  Consequently, it is not possible to generate the baryon
asymmetry of the universe through leptogenesis in supersymmetric
models where $m_N>m_{W_\R}$, or when $m_{W_\R}\lesssim10^7\gev$.

\begin{center}
{ACKNOWLEDGEMENT}
\end{center}

We acknowledge the hospitality of the DESY Theory Group and one of us
(US) also acknowledges financial support from the Alexander von
Humboldt Foundation.  The work of EM was supported in part by the 
U.~S.~Department of Energy under Grant No.~DE-FG03-94ER40837.

\bibliographystyle{unsrt}

\end{document}